\documentclass[fleqn]{elsarticle}

\usepackage{subfigure}
\usepackage{graphicx}
\usepackage{color}
\usepackage{amssymb,amsmath}
\usepackage{wasysym}

\newcommand{\tab}{\hspace*{2em}}
\newcommand{\ltapprox}{\lower 2pt \hbox{$\, \buildrel {\scriptstyle <}\over {\scriptstyle \sim}\,$}}
\renewcommand{\gtrapprox}{\lower 2pt \hbox{$\, \buildrel {\scriptstyle >}\over {\scriptstyle \sim}\,$}}

\newcounter{bla}

\journal{Computer Physics Communications}
\begin{document}
\begin{frontmatter}

\title{\textsc{tau}: A 1D radiative transfer code for transmission spectroscopy of extrasolar planet atmospheres}

\author{M.~D.~J.~Hollis}
\ead{mdjh@star.ucl.ac.uk}

\author{M.~Tessenyi}
\author{G.~Tinetti}

\address[ucl]{Department of Physics and Astronomy, University College London, Gower Street, London WC1E 6BT, UK}

\begin{abstract}
\label{sec:abstract}
	The \textsc{tau} code is a $1D$ line-by-line radiative transfer code, which is generally applicable for modelling transmission spectra of close-in extrasolar planets. The inputs are the assumed pressure-temperature profile of the planetary atmosphere, the continuum absorption coefficients and the absorption cross-sections for the trace molecular absorbers present in the model, as well as the fundamental system parameters taken from the published literature. The program then calculates the optical path through the planetary atmosphere of the radiation from the host star, and quantifies the absorption due to the modelled composition in a transmission spectrum of transit depth as a function of wavelength. The code is written in C++, parallelised using OpenMP, and is available for public download and use from \textit{http://www.ucl.ac.uk/exoplanets/}.
\end{abstract}


\end{frontmatter}

\section{Introduction}
\label{sec:intro}
	The discovery of the first extra-solar planet by~\citet{Wolszczan+92} in 1992 marked the beginning of an entirely new area of astronomy, and the detection of transiting exoplanets by~\citet{Mayor+95} in 1995 for the first time made it possible to attempt to characterise planets outside of our own Solar System. The subsequent rapid advances in the field have led to a large number of currently confirmed planets, predominantly so-called `hot Jupiters', gas giants orbiting very close in to their parent stars. Such planets exist in extreme regimes in terms of temperatures, stellar fluxes and tidal forces, and bridge the gap between planetary and stellar bodies; hence a wide variety of models (see for example~\citet{Showman+02,Cooper+06,Venot+12,Irwin+08,Fortney+08,Seager+07,Burrows+06}) have been developed to attempt to understand the orbital and atmospheric dynamics and chemistry of these intriguing objects, and to retrieve the bulk and atmospheric compositions and thermal profiles. Such models can also be extended to understanding the atmospheres of terrestrial planets, with obvious applicability to the search for habitable worlds. Though data for such objects are limited at the present time, observations will be forthcoming, making use of the capabilities afforded by major future space missions such as the \textit{James Webb Space Telescope} (JWST) and the \textit{Exoplanet Characterisation Observatory} (EChO). 

The \textsc{tau} code is a $1D$ radiative transfer code for transmission spectroscopy at infra-red wavelengths of gaseous and terrestrial exoplanets, used for example in characterising close-in hot Jupiters~\citep{Tinetti+07}. It uses a line-by-line integration scheme to model transmission of the radiation from a parent star through the atmosphere of an orbiting planet, a so-called `forward model', equating physically to observations of the radius ratio as a function of wavelength in primary transit geometry. This allows the user to estimate the abundances of absorbing molecules in the atmosphere, by running the code for a variety of hypothesised compositions and comparing to any available observations. 

Specifically, the algorithm calculates the optical depth of the planetary atmosphere at a particular wavelength, with a hypothesised (model) bulk composition and trace molecular abundances, and given the atmospheric structure and absorbing behaviour of those molecules. Rayleigh scattering in the bulk atmosphere is calculated for all of the specified bulk constituents, and the optical depths due to this and the trace molecular absorption are then used within the geometry of the system to calculate an effective radius of the planet plus atmosphere (i.e. the conventional radius modified by the atmospheric absorption). A transit depth can hence be calculated as the ratio of the squared radii of the planet and the star, and the process repeated for every wavelength in the required spectral range, to build up a spectrum showing absorption as a function of wavelength. 

Required input files to the code are a pressure-temperature (`P-T') profile and absorption cross-sections as a function of wavelength for the species hypothesised to be present in the atmosphere. Absorption cross-sections are generally available from external sources (see Section~\ref{ssec:inputs}), and for the profile and other optional inputs, the sample files provided can be altered as required, or generated anew by other means. The stellar radius as a function of wavelength can either be assumed constant (default) or given as an optional input to the code, as can collision-induced absorption coefficients. Sample files for all required inputs are provided in the tar ball, downloadable from \textit{http://www.ucl.ac.uk/exoplanets/}. 

The structure of this paper is as follows: Section~\ref{sec:theory} covers some of the necessary theory, and Section~\ref{sec:install} briefly describes how to install the program before Section~\ref{sec:implement} discusses in detail the inputs, outputs, run modes and program structure and algorithms. Finally, Section~\ref{sec:application} illustrates how to use the model, in this case for some of the molecules and scatterers relevant to near-infrared (`NIR') transmission spectroscopy of extrasolar giant planets.

\section{Background Theory}
\label{sec:theory}

\subsection{System geometry}
\label{ssec:geometry}
\begin{figure}[h!]
\caption{Diagram illustrating the geometry of a primary transit, with light from the star blocked as the planet moves around in its orbit and passes between star and observer.}
\label{fig:transit}
\smallskip
\scriptsize
\centering
	\includegraphics[width=0.9\columnwidth]{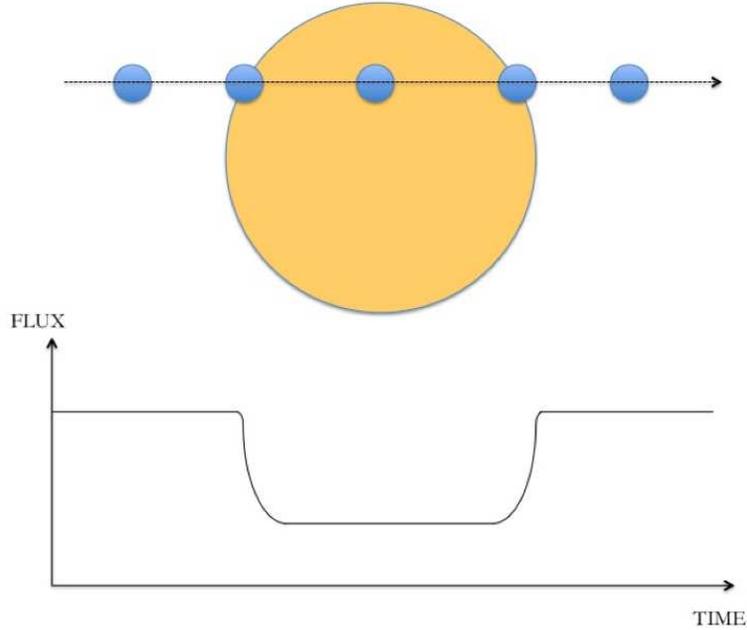}
\end{figure}

	The key to being able to characterise exoplanets and infer vital information such as the atmospheric composition is the observation of planetary transits (e.g.~\citet{Seager+00, Brown01,Tinetti+07}). In the case of a primary transit (the planet passing between the observer and the star, as shown in Figure~\ref{fig:transit}), light from the star is blocked by the planet, producing a reduction in the observed flux. This reduction, or the \textit{transit depth}, can be expressed simply as the ratio between the radius of the planet, $R_p$ and the radius of the star, $R_*$, given by,

\begin{equation}
\label{eq:tran_depth}
transit\,depth \,=\, \frac{R_{p}^{2}}{R_{*}^{2}}. 
\end{equation}

Transmission spectroscopy, as the name suggests, deals with the transmission of the stellar light through the planetary atmosphere, and hence applies in this primary transit scenario. In particular, it probes the \textit{terminator} region, the part of the planet separating the dayside and nightside. When the planet passes in front of the star, the light is filtered through this section of the atmosphere of the planet in order to reach the Earth, introducing a very small, but nonetheless measurable, variation in the observed flux. Since different molecules absorb some of the stellar light at different wavelengths, by measuring the spectrum (the observed flux as a function of wavelength) of the planet relative to the star, the various absorption features can be identified, and the types and amounts of molecules present in the atmosphere of the planet can be inferred. 

The \textsc{tau} code enables this kind of study, and can be used to generate model spectra of both gaseous and terrestrial exoplanet atmospheres to be compared with observed data (e.g. by minimising some goodness-of-fit parameter) in order to infer compositions (see for example~\citet{Waldmann+13,Danielski+13}).

\subsection{Radiative transfer}
\label{ssec:rad_trans}
	It is instructive to consider a simple absorption scheme for radiation passing through a gas, e.g. a planetary atmosphere. If the intensity at the start of an arbitrary path is $I_{\lambda}$, then as the radiation passes through the medium along a path $ds$ it is weakened by absorption and scattering by the material, with the extinction effects expressed as,

\begin{equation}
\label{eq:path_extinc}
dI_{\lambda}\,=\, -\, I_{\lambda}\, \sigma_{\lambda}\, \rho\, ds,
\end{equation}

where $\rho$ is the (mass) density of the material in $kg\,m^{-3}$ and $\sigma_{\lambda}$ is the \textit{mass extinction cross section} in $m^{2}\,kg^{-1}$ for radiation of wavelength $\lambda$, or the sum of the absorption and scattering cross sections. 

This can be rearranged to give,

\begin{equation}
\label{eq:path_extinc2}
\frac{dI_{\lambda}}{\sigma_{\lambda}\, \rho\, ds}\,=\, -\, I_{\lambda},
\end{equation}

from which the quantities \textit{optical path length}, $u$, and \textit{optical depth}, $\tau$, can be defined as,

\begin{equation}
\label{eq:opt_path}
u \,=\, \int \rho\, ds,
\end{equation}

\begin{equation}
\label{eq:tau}
\tau_{\lambda} \,=\, \sigma_{\lambda}\, \int \rho\, ds \,=\, \sigma_{\lambda}\, u,
\end{equation}

respectively.

It follows that the intensity at a point at altitude $z$ is given by integrating along the vertical path from the top of the atmosphere $z_{\infty}$ as,

\begin{equation}
\label{eq:intensity}
I_{\lambda}(z)\,=\,I_{\lambda}(0)\, exp\,\left( - \sigma_{\lambda}\, \int_{z}^{z_{\infty}}\, \rho\, dz \right),
\end{equation}

which is the monochromatic intensity at altitude $z$ as given by the \textit{Beer-Bouguer-Lambert Law},
		
\begin{equation}
\label{eq:bbl}
I_{\lambda}(z) \,=\, I_{\lambda}(0)\, e^{-\tau_{\lambda}(z)},
\end{equation}

quantifying the absorption of radiation entering from above and passing down vertically through the atmosphere. For more information on radiative transfer in planetary atmospheres, the reader is directed to texts such as~\citet{Liou02}.

\subsection{Absorption cross-sections}
\label{ssec:abs_cs}
	The distance travelled through the atmosphere is hence an important quantity in determining the transfer of radiation down through the atmosphere. This in turn depends on various parameters, and fundamentally on the extinction cross-section, $\sigma_{\lambda}$ in Equation~\ref{eq:path_extinc}. Neglecting scattering for simplicity, this becomes the absorption cross-section, which is calculated in turn from the \textit{absorption coefficient}, $\alpha(\lambda)$. This is a quantity calculated by considering the quantum behaviour of the molecules and how they absorb and emit radiation - for example different molecules have different electronic transitions, and both rotate and vibrate spatially in many different orientations and configurations at various characteristic frequencies (wavelengths), i.e. with different energies. The possible transitions between these energy levels produce many absorption and emission lines for a given molecule, many of which may occur at the same wavelengths but represent different motions and transitions between states with different quantum numbers. Hence a \textit{line list} must be produced, which provides each emission and absorption line intensity as a function of wavelength, for each different transition between all quantum states, for a given molecule. 
	This line list can then be used, in conjunction with the partition function for that molecule and correct line shape (i.e. considering the necessary line broadening effects) at the correct temperature and pressure, in order to combine all transitions over wavelength. This gives the variation of absorption coefficient with wavelength for a given molecule at a given temperature, with the `strength' of the absorption quantified in the value of the absorption coefficient for that particular wavelength. These absorption coefficients are normally expressed in units giving the absorption per unit mass, or attenuation per unit length and density (relative to stp), and can be converted, depending on their units, to mass absorption cross-sections, $\sigma_{\lambda}$, with units $cm^{2} g^{-1}$. Finally, removing any mass or density dependence by converting these to units of $cm^{2}$ gives the absorption cross-sections, $\sigma(\lambda)$, as a function of wavelength as required for this code.

\subsection{Optical depth}
\label{ssec:opt_depth}
\begin{figure}[h!]
\caption{Geometry of a primary transit observation, illustrating the paths of the stellar photons filtered through the planetary atmosphere.}
\label{fig:geometry}
\smallskip
\scriptsize
\centering
	\includegraphics[width=0.9\columnwidth]{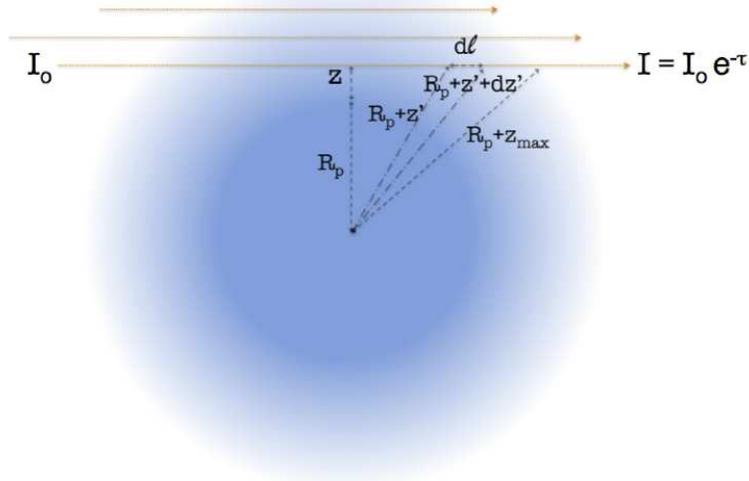}
\end{figure}

	The optical path of the stellar photons is shown in Figure~\ref{fig:geometry}. The optical path is calculated according to the geometry of this transverse view through the terminator of the transit system, and also requires a knowledge of the quantity of each molecule $i$ in the path, expressed by the dimensionless \textit{mixing ratio}, $\chi_i$. This is the ratio of the amount of that molecule to the amounts of all other molecules combined, i.e. it is a fractional abundance or amount of the gas in question relative to the rest of the atmosphere. Hence the concentration of a particular molecule $i$ with a number density $\rho_{N}\,[m^{-3}]$ is $\chi_i\,\rho_{N}$, and Equation~\ref{eq:tau} can then be extended to give the optical depth at an altitude $z\,[m]$ for a given molecule $i$ with cross-section $\sigma_{i}\,[m^2]$ as, 

\begin{equation}
\label{eq:tau_i}
\tau_{i}(\lambda,z) \,=\, 2\, \int_{0}^{l(z)}\, \sigma_{i}(\lambda)\, \chi_{i}(z')\, \rho_{N}(z')\, dl.
\end{equation}

\noindent The total depth for all molecules for that path is then,

\begin{equation}
\label{eq:tau_tot}
\tau(\lambda,z) \,=\, \sum_{i=1}^{N}\, \tau_{i}(\lambda,z),
\end{equation}

the sum of the contributions from the $N$ individual molecules present in the atmosphere. This is finally converted to an equivalent atmospheric depth $A(\lambda)$ by summing all of the viewing paths as,

\begin{equation}
\label{eq:Ratm}
A(\lambda) \,=\, 2\, \int_{0}^{z_{max}}\, (R_{p}\,+\,z)\,(1\,-\,e^{-\,\tau(\lambda,z)})\,dz,
\end{equation}

with the total transit depth at that wavelength $D(\lambda)$ found by,

\begin{equation}
\label{eq:tran_depth2}
D(\lambda) \,=\, \frac{R_{p}^{2}\, +\,A(\lambda)}{R_{*}^{2}}. 
\end{equation}

A model of the actual stellar spectrum is thus not required at any point. Absorption by the molecules present in the atmosphere reduces the overall final transmitted flux at wavelength $\lambda$ by the factor $(1\,-\,e^{-\,\tau(\lambda,z)})$, which is the same effect as having a totally opaque body with a slightly larger radius transiting the star. The absorption can therefore be quantified by a simple radius ratio (i.e. a conventional transit depth) at each wavelength, and a spectrum can hence be constructed showing the absorption as a function of wavelength for the input model parameters.

\section{Installation}
\label{sec:install}

\begin{itemize}
\item{\textbf{Step 1:} Download the tar ball \verb+tau.tar.gz+ from \textit{http://www.ucl.ac.uk/exoplanets/}.}

\item{\textbf{Step 2:} In the directory where the file is downloaded, type into the terminal: \\ \\ \verb+tar -xvf tau.tar.gz+ \\ \\ This will extract all of the source files into a directory named \verb+tau+.}

\item{\textbf{Step 3:} To compile and make the executable, whilst in the \verb+tau+ directory type: \\ \\ \verb=g++ -lgomp -fopenmp -o tau ./src/tau.cpp= \\ \\ This creates the executable file \verb+tau+ in the current directory. The header file containing the required functions and scientific constants is located in the directory \verb+./tau/src/+, output files are placed in the directory \verb+./tau/out/+, and the required input files should be in the directory \verb+./tau/run/+. Documentation for the code is in \verb+./tau/doc/+.}

Note that for the compilation, it is assumed that g++ from GCC, the Gnu Compiler Collection, is present on the system. If not, it may be downloaded from \textit{http://gcc.gnu.org}. Additionally, the library \verb+omp.h+ from the OpenMP API specification for parallel programming must be installed, obtainable from \textit{http://openmp.org/}.
\end{itemize}

\section{Implementation}
\label{sec:implement}
	The \textsc{tau} code is a line-by-line radiative transfer code following~\citet{Ehrenreich+06, Tinetti+07, Tinetti+12}, which calculates the transmittance of radiation through an atmosphere given the temperature structure (variation of temperature with pressure/altitude), the assumed gaseous constituent abundances, or mixing ratios (either hard-coded in pre-compilation, or read in from file, depending on the run mode), and a defined spectral wavelength grid on which to calculate the results (with resolution depending on how much detail is required in the final model spectrum). 

The code assumes \textit{local thermodynamic equilibrium} (`LTE') conditions for an input atmospheric profile, which most commonly also assumes \textit{hydrostatic equilibrium}. The atmosphere itself is divided into horizontal layers, which are assumed to be sufficiently thin as to be regarded as homogenous within each layer. Each layer is hence uniquely defined by the pressure/altitude and temperature value, determined according to the precise nature of the P-T profile input by the user (i.e. the existence and exact form of any temperature inversions, or the assumption of a non-isothermal atmosphere). 

The optical path length in each layer can hence be calculated according to the geometry of the system, and the optical depth follows by a simple product between this length, the abundance of each type of absorber present, and the absorption cross-section for each absorber. For each layer-segment along the viewing path then, the separately calculated contributions of all species to the optical depth may be added together, and these optical depths may themselves be added together to provide a value integrated along the entire viewing path to the observer. Rayleigh scattering within the atmosphere, and optionally opacity due to collision-induced absorption, are also included, but at present a comprehensive treatment of cloud scattering is omitted, for simplicity. 

The code is written in C++, with one file containing the main program (\verb+./tau/src/tau.cpp+) and one file containing the libraries, some scientific constants and the external functions called in the main code (\verb+./tau/src/functions.h+). It includes some functions from the OpenMP parallel programming library - these are used to retrieve the total program runtime, and for some calculations in the interpolation functions. These calculations will by default be performed on the maximum number of cores available; if only a single core is available, the program should be compiled as normal and the program will effectively run in serial mode. The speedup factor of the parallel implementation depends not only on the number of cores used, but also various input factors, such as the number of absorbers and the sizes and wavelength ranges of their input cross-section files, and the required resolution of the output spectrum. The main processing bottleneck is the interpolation of the absorption cross-section files to the wavelength grid (see Section~\ref{ssec:structure}).

\subsection{Input files}
\label{ssec:inputs}
	The \textsc{tau} code requires the following files to run:
\begin{itemize}
\item{atmospheric P-T profile}
\item{absorption cross-section as a function of wavelength for each modelled absorber/trace molecule}
\item{[optional] stellar radius $R^{*}(\lambda)$}
\item{[optional] CIA absorption coefficients $\alpha_{cia}(\lambda)$}
\item{[optional] cloud mass absorption coefficients $\alpha^{'}_{cld}(\lambda)$}
\end{itemize}

\noindent All of these files should be stored in the directory \verb+./tau/run/+, and are read in at various points of the code, as detailed in Section~\ref{ssec:structure}. Note the formats and units of the input files must be adhered to - if the absorption cross-section data to be used are originally expressed in wavenumbers, they must be converted before being read in by the code (for example using the relation $\lambda\,=\,(1 \times 10^{4})  / \nu$, with the wavelength $\lambda$ in $\mu m$ and the wavenumber $\nu$ in $cm^{-1}$).

\subsubsection{P-T profile}
\label{sssec:profile}
The atmospheric profile file (see sample file \verb+./tau/run/profile.atm+) describes the temperature and pressure variation with altitude in a given number of horizontal layers from the top to the bottom of the model atmosphere. It must be in (at least) 4-column format, separated by a non-zero amount of whitespace, containing columns of pressure ($p$, in $Pa$), temperature ($T$, in $K$) and altitude ($z$, in $km$) in that order, with then at least one column of mixing ratios ($\chi$, dimensionless) as a function of altitude, with one column per intended molecular absorber to be modelled in the atmosphere. This fourth column must still be present even if Run Mode 9 is selected, and the values are then overridden by the value hard-coded in (see Section~\ref{ssec:modes}). The columns must start with the uppermost (i.e. lowest pressure) atmospheric layer, and the number of atmospheric layers is $39$ in the sample file, but this may be any value the user chooses. In addition, a header of 10 lines, the contents of which is ignored by the program, must be present at the start of the file. The file is read in by the function \\ \\ \verb+readAtmFile_H+ \\ \\ and associated sub-functions, the source code for which is located in the header file. The atmospheric profile can be calculated in whatever manner and using whichever program the user chooses, so long as the final files adhere to the specified format. 

\subsubsection{Absorption cross-sections}
\label{sssec:csection}
The molecular absorber cross-section files (see sample files \\ \verb+./tau/run/h2o_T.abs+\footnote{calculated for water at temperature $T$ from the BT$2$~\citep{Barber+06} line list.}) give the variation in absorption cross-section as a function of wavelength, with one file per molecule/trace absorber present in the modelled atmosphere. They are in a 2-column format, separated by a non-zero amount of whitespace, with the first column being the wavelength ($\lambda$, in $\mu m$) and the second being the absorption cross-section at that wavelength ($\sigma(\lambda)$, in $cm^{2}$), in that order. There is no header associated with these files, and the files may either be in wavelength-decreasing or wavelength-increasing order, with any re-ordering, if necessary, being performed within the program. The files are read in by the function \\ \\ \verb+readAbsFile_H+ \\ \\ and associated sub-functions, the source code for which is located in the header file. Absorption cross-sections may be obtained from various external sources, such as the \textsc{hitran}~\citep{Rothman+09} and \textit{Exomol}~\citep{Tennyson+12} molecular line list projects, and again, may have to be converted to the required format and units before use with this code. Care must be taken to use the appropriate line lists for the modelled atmospheric conditions, since there are strong dependencies of absorption coefficients, and hence cross-sections, on factors such as temperature and pressure. A detailed study of this temperature dependence and the calculation of molecular line lists is however outside the scope of this document, and the reader is directed to~\citet{Tennyson+12} for further information. 

\subsubsection{Stellar radius}
\label{sssec:radius}
A changing stellar radius as a function of wavelength may also be modelled, controlled, as for the CIA and cloud optional inputs, by the state of the \verb$sw_$ switches (e.g. \verb$sw_rad$ for the stellar radius input). The file (e.g. \verb+./tau/run/rad_star.rad+) must be in 2-column format, separated by a non-zero amount of whitespace, containing columns of wavelength ($\lambda$, in $\mu m$) and stellar radius ($R^{*}$, in $m$) in that order, in wavelength-decreasing order. No header should be present at the start of the file, which is read in and interpolated to the model grid by the function \\ \\ \verb+interpolateCS_H+ \\ \\ the source code for which is located in the header file. The stellar radius values are necessary at all wavelengths to calculate absorption at those wavelengths. If \verb$sw_rad$ is set but the range of the stellar radius input file data does not span the wavelength range set for the model wavelength grid, a warning message will be displayed. Wavelengths for which no data are provided will then be set to the constant stellar radius value defined with the other system parameters at the head of the code. This may lead to inaccurate results, hence care should be taken that the stellar radius as a function of wavelength is known across the entire desired wavelength range of the transmission spectrum calculation if \verb$sw_rad$ is set. Also note that the value of \verb$s_rad_fac$ will be overridden in this case and will have no effect.

\subsubsection{CIA coefficients}
\label{sssec:cia}
The file containing the collision-induced absorption (`CIA') coefficients (e.g. \verb+./tau/run/h2_h2_1000K.cia+) gives the variation in absorption coefficient as a function of wavelength. They are in a 2-column format, separated by a non-zero amount of whitespace, with the first column being the wavelength ($\lambda$, in $\mu m$) and the second being the absorption coefficient at that wavelength ($\alpha_{cia,\lambda}$, in units of $cm^{-1} amagat^{-2}$). There is no header associated with these files, and they must be in wavelength-decreasing order. The files are read in by the function \\ \\ \verb+interpolateCS_H+ \\ \\ the source code for which is located in the header file. CIA coefficients may be obtained from external sources such as~\citet{Borysow+97,Borysow+01} in the units stated above (and converted within the program), or from~\citet{Richard+12} in different units (requiring additional conversion). If the input data do not span the wavelength range of the model grid, the optical depths will be set to zero for the relevant wavelengths, and hence there will be no absorptive effect on the spectrum here. If extra CIA is desired (e.g. from H$_{2}$-He collisions), this may be added `artificially' by using the CIA file as another trace molecular absorber, with the coefficients converted to cross-sections as per the format requirements and specifying a mixing ratio of $\chi=1$. More details regarding the calculation can be found in~\ref{sec:app_scatter}.

\subsubsection{Cloud coefficients}
\label{sssec:cloud}
Cloud effects are not modelled in this version of the code for simplicity, although they will be treated in detail in future versions. However, if scattering data are available from external studies of cloud physics, they may be coupled to the code and input as files (e.g. \verb+./tau/run/cloud1.cld+). Since the Lorenz-Mie theory of particle scattering is complex and dependent on the properties of the clouds (e.g. particle shape and size), calculation of these parameters must be performed externally, and any optional input files used must contain the cloud mass absorption coefficient as a function of wavelength, for each different cloud type/composition. The files must be in 2-column format, separated by a non-zero amount of whitespace, containing columns of wavelength ($\lambda$, in $\mu m$) and cloud mass absorption coefficient ($\alpha^{'}_{cld}(\lambda)$, in $cm^{2} g^{-1}$) in that order, in wavelength-decreasing order. No header should be present at the start of the files, which are read in and interpolated to the model grid by the function \\ \\ \verb+interpolateCS_H+ \\ \\ the source code for which is located in the header file. If the input data do not span the model wavelength range, the missing optical depths will again be set to zero (i.e. will have no effect on the spectrum). If cloud opacity data are to be input, the vertical extent of the clouds must also be set, by specifying the maximum (i.e. the cloud base) and minimum pressure levels at which the clouds exist. These values are currently set to be the same for every different input cloud file for simplicity, since future versions of the code will treat cloud effects more completely, and will have the capability of specifying a different vertical extent for each type of cloud. The default values model clouds extending through atmospheric layers from $0.1\,bar$ to $1.0\,mbar$.

\subsection{Output files}
\label{ssec:output}
	The output from the code is an ascii file called \verb+tau_output.dat+, which gives the variation in radius ratio between the planet and the star as a function of wavelength, as per Equation~\ref{eq:tran_depth2}. This file is in two columns, the first containing the wavelength ($\lambda$, in $\mu m$) and the second containing the amount of absorption ($D(\lambda)$, a dimensionless ratio). From Equation~\ref{eq:tran_depth2} it can hence be seen that a larger value for the absorption is equivalent to a larger transit depth, and hence a larger equivalent planetary radius, at that wavelength. This can then be plotted to give the full absorption spectrum of the specified molecular absorbers in the given atmosphere on the planet in primary transit.

\subsection{Run modes}
\label{ssec:modes}
	The current version of the code has the option to run in 3 modes, as detailed below. Run Mode 0 is used to calculate the spectrum for a single trace molecular absorber, Run Mode 1 is used to model multiple molecular absorbers, and Run Mode 9 is used for a quick run with custom settings. Run Modes 0 and 1 allow the \verb+.atm+ and/or \verb+.abs+ file inputs to be specified as command line arguments or interactively, and Run Mode 9 provides a quick testing mode, allowing all interactive elements to be bypassed and all input files and mixing ratio values hard-coded prior to compilation. The usage instructions and mode formats can be displayed by typing \verb+./tau+ at the command line, and the option sorting and default file definitions can be found in the function \\ \\ \verb+optionSort_H+ \\ \\ in the header file.

\begin{itemize}
\item{\textbf{Run Mode 0:} The run command for this mode is \verb+./tau 0 [[atmfile]] [absfile]+
\\ \\
This mode allows the user to specify the atmospheric profile file and a single absorption cross-section file (i.e. to model a single absorber molecule in the atmosphere). If run in this mode, both files are optional command line arguments, and if nothing is entered for either the code reverts to the default files, \verb+./run/profile.atm+ and \verb+./run/h2o_1500K.abs+. It is possible to only enter a filename for the atmospheric profile, and leave the code to take the default absorption cross-section file, but if the \verb+.abs+ file is to be entered, so the \verb+.atm+ file must also be. Mixing ratios are read in from the atmospheric profile file.}

\item{\textbf{Run Mode 1:} The run command for this mode is \verb+./tau 1 [atmfile]+
\\ \\
This mode allows the user to specify on the command line the atmospheric profile file and (interactively) multiple absorption cross-section files (i.e. to model an arbitrary number of absorber molecules in the atmosphere). If run in this mode, the file is an optional command line argument, and if nothing is entered the code reverts to the default file, \verb+./run/profile.atm+. Mixing ratios are read in from the atmospheric profile file (one column per absorption cross-section file).}

\item{\textbf{Run Mode 9:} The run command for this mode is \verb+./tau 9+	
\\ \\
This mode allows the user to bypass all interactive input, and to specify all input files and mixing ratios in the program code pre-compilation. This mode is useful for example when comparing models where the atmospheric constituents and most of the mixing ratios remain constant between runs but one or a few are altered. The default atmospheric profile and absorption coefficient files are set in the file \verb$functions.h$, and mixing ratios are set to $\chi = 10^{-5}$ in the file \verb$tau.cpp$.}
\end{itemize}

\subsection{Run parameters}
\label{ssec:run_params}
	The \textsc{tau} code requires certain parameters to be specified in order to produce the required model, as follows.

\begin{itemize}
\item{Parameters defined by user:}

\verb+rad_fac+, \verb+s_rad_fac+ \\ \tab variables to increase or decrease the assumed planetary radius. These radius modifiers can be altered as required to vary the planetary and stellar radii by these percentages, in order improve the fit to any observations. The variables represent the, sometimes significant, uncertainties in estimations of the planetary and stellar radii respectively, deriving from uncertainties in stellar models and the ill-defined planetary `surface' for gas giants (the altitude (generally unknown) at which the atmospheric pressure is equal to $1$ or $10\,bar$). \\

\verb+lambda_min+, \verb+lambda_max+ \\ \tab the desired minimum and maximum wavelengths of the output spectrum, in $\mu m$. If these are set such that $\verb+lambda_max+ < \verb+lambda_min+$, the program will revert to automatically adopt the global minimum and maximum wavelength values from the input absorption cross-section files as the bounds of the calculated spectrum, and use the default wavelength resolution. \\

\verb+lambda_res+ \\ \tab the default resolution $\Delta\lambda$ of the output spectrum, where spectral resolving power $R$ at a wavelength $\lambda$ is defined by $R\,\equiv\,\lambda\,/\,\Delta\lambda$. If the wavelength range is defined manually (i.e. $\verb+lambda_max+ > \verb+lambda_min+$) then the resolution can be automatically updated to be one two-thousandth of the total range. \\

\verb+sw_rad+, \verb+sw_cia+, \verb+sw_cld+\\ \tab switches to include (if not set to $0$) a stellar radius varying with wavelength, H$_{2}$-H$_{2}$ CIA and extra opacities due to clouds respectively, which can be turned off if required, but CIA should in general be left on. H$_{2}$ Rayleigh scattering is always included, and becomes increasingly important at smaller wavelengths (since the scattering cross-section $\sigma_{R}\,\sim\,\lambda^{-4}$), and CIA provides a base absorption floor at pressure levels in the atmosphere of greater than $p\,\approx\,1\,bar$. More details regarding the expressions used for the extra scattering effects can be found in~\ref{sec:app_scatter}. The cloud switch \verb$sw_cld$ should be set to equal the number of different clouds for which input files are supplied. \\

\verb+mixdef+ \\ \tab the default mixing ratio, taken as the mixing ratio for all trace molecules at all heights, for Run Mode 9 only.
\newline \newline

\item{Parameters taken from literature measurements/estimations:}
\\ \\ \verb+Rp+ \\ \tab the radius of the planet to be modelled, in $R_{\jupiter}$.
\\ \\ \verb+Rstar+ \\ \tab the radius of the host star, in $R_{\tiny{\astrosun}}$.
\\ \\ \verb+semimajor+ \\ \tab the semi-major axis of the planetary orbit, in $AU$.
\\ \\ \verb+grav+ \\ \tab the gravitational acceleration at the planetary surface, in $m s^{-2}$.
\\ \\ \verb+temp+ \\ \tab the planetary temperature, in $K$.
\newline \newline

\item{Rayleigh scattering parameters:}
\\ \\ Different bulk compositions are accounted for using the class \verb+Mol+. Each constituent species of the bulk atmosphere is specified as a variable of type \verb+Mol+, defined for example by \verb+H2("H2",2.0,2.0e-9,1.0001384)+, where the respective parameters are the name of the molecule, the relative molecular weight (in $u$), the particle radius (in $m$), and the refractive index (assumed constant with wavelength). A class method is then used to add each species to the \verb+Atmos+ class with a specified fractional abundance (N.B. the total sum of the abundances must equal $1$, to make up the entire bulk atmosphere), and the mean molecular weight of the entire atmosphere is calculated. This, along with the surface gravity and pressure used to create the atmospheric P-T profile, defines the scale height and determines whether the planet is terrestrial in nature, or a hydrogen-dominated giant.
\newline \newline

\item{Filenames specified by user:}
\\ \\ \verb+rad_file+ \\ \tab file containing the stellar radii values $R^{*}(\lambda)$.
\\ \\ \verb+cia_file+ \\ \tab file containing CIA coefficients $\alpha_{cia}(\lambda)$.
\\ \\ \verb+cld_file+ \\ \tab array of files containing cloud optical depth values $\tau_{cld,i}(\lambda)$.
\\ \\ \verb+arg_outfile+ \\ \tab output file to write the final spectrum to.
\newline
\end{itemize}

\subsection{Program structure}
\label{ssec:structure}
	The code is run by typing \\ \\ \verb+./tau mode [input files]+ \\ \\ depending on the run mode required (see Section~\ref{ssec:modes}). It begins by including the required C++ libraries and scientific constants, along with the prototypes for the external functions called in the main code, located in the file \\ \\ \verb+./tau/src/functions.h+ \\ \\ The run parameters are then initialised and defined, and some option sorting and memory allocation is performed.

The next main stage is to read in the absorption cross-sections for all of the modelled absorber molecules (see Section~\ref{ssec:inputs}). A two-dimensional vector container, \verb+sigma_array+, is first declared, and resized such that one `row' is used to store the cross-sections as a function of wavelength for each molecule, with the top `row' in the object reserved for the output wavelength grid. In order to maintain a well-defined spectral resolution in the final wavelength grid, the data from each absorption cross-section file are read into temporary containers, and then cross-interpolated in the function \\ \\ \verb+interpolateAbs_H+ \\ \\ to the same wavelength grid, defined using the minimum and maximum wavelength values (either input by the user, or taken automatically from all \verb+.abs+ files) and with spacing given by the user-defined parameter \verb+lambda_res+. If any absorption cross-section files do not cover the same wavelength range as that determined by the \verb+lambda_min+ and \verb+lambda_max+ values, the absorption cross-sections for those molecules are set to zero in these regions. Hence those molecules will have no effect on the spectrum throughout wavelength ranges for which they have no input data. The interpolated data are finally stored in the \verb+sigma_array+.

Vector containers are then created to hold data from any optional file inputs, with radius (\verb$rad_star$) and CIA (\verb$cia_coeffs$) data in one-dimensional vectors, and cloud data (\verb$cld_tau$) in a two-dimensional vector, with optical depths as a function of wavelength for each different cloud type/condition held in a different `row' in this object. Depending on the states of the optional input switches, the function \\ \\ \verb+interpolateCS_H+ \\ \\ may then be called to also interpolate the data to the new wavelength grid. If the switches are off, then the vectors are initialised to their respective defined constant values, as detailed in Section~\ref{ssec:inputs}, regardless of the contents of the specified files, or even if the specified files exist. If cloud opacity data is to be input, the cloud vertical extent is also defined here by specifying the pressure levels of the cloud base and top. 

The atmospheric profile data are then read in by the code, with each atmospheric state parameter defined as an array of floats, with size equal to the number of vertical layers as read from the file, such that each atmospheric layer is then uniquely defined by the values of the pressure, temperature and altitude variables. The mixing ratios for each modelled absorber gas are read in similarly, or in the testing run mode can be hard-coded in here. The atmospheric state parameters are then also used to calculate the number density in each vertical layer, \verb+rho[layer]+, from the standard thermodynamic state and ideal gas equations, which can be arranged\footnote{Note that using this form of the equations eliminates the need to calculate the conventional density $\rho$ (mass density), and hence the volume $V$ of any layer in the atmosphere.} to give,

\begin{equation}
\label{eq:num_dens}
\rho_{N}[layer] \,=\, \frac{p}{k_{B}\,T},
\end{equation}

where $k_{B}$ is Boltzmann's constant, and $p$ and $T$ are the pressure and temperature respectively for each layer. 

The final section of the code calculates the path length integral given by Equation~\ref{eq:tau_i} for each molecule, summing these and converting the value to the equivalent transit depth as a function of wavelength. The program declares an array to hold the final absorption values, one per wavelength value, and then loops over wavelength and the number of vertical layers, to first calculate the geometric path length \verb+dl+ at altitude \verb=z[k+j]=, and hence the optical path length (the product of the geometric path length and the number density of the layers along the path). The optical path at each altitude \verb=tau[j]= is then found for each absorber/scatter by taking the product of the optical path length with the abundance of that constituent. Calculations of the Rayleigh scattering, CIA and (estimated) cloud contributions to the opacity are also performed at this point in the code, since they are function of parameters that vary in each layer of the optical path. The calculation procedure for these extra opacities is detailed in~\ref{sec:app_scatter}. 

Finally the equivalent radius of the planet (plus atmosphere) for all viewing paths (i.e. integrated over altitude) as a function of wavelength can then be calculated as per Equations~\ref{eq:Ratm} and \ref{eq:tran_depth2}. Note that the equations in the code calculate the half values for the quantities \verb+dl+ and \verb+integral+, which can then simply be doubled due to the symmetry in Figure~\ref{fig:geometry} in order to find the full values. The specified output file is then opened and the model data written to it, and the code finishes with a call to a library to calculate the total runtime.

\subsection{Code validation}
\label{ssec:validation}
	In order to validate the output from the code, a comparison to an analytic solution may be performed. As discussed in~\citet{desEtangs+08, Howe+12}, for opacity sources that are independent of temperature and pressure and proportional to the wavelength the cross-section may be expressed in the form of a scaling law,
\begin{equation}
\label{eq:sig_scale}
	\sigma = \sigma_{0} \left(\frac{\lambda}{\lambda_{0}}\right)^{\alpha}  .
\end{equation}

The slope of the observed planetary radius as a function of the logarithm of the wavelength should then be a straight line with gradient $\alpha$, 
\begin{equation}
\label{eq:LdE}
	\frac{d R_{p}}{d ln \lambda} = \alpha H  ,
\end{equation}
with $H$ being the atmospheric scale height, $H\,\equiv\,k_{B}T/\mu g$. If an opacity source is known to follow Equation~\ref{eq:sig_scale} and if the equilibrium temperature is known, Equation~\ref{eq:LdE} can hence be used to calculate the mean molecular weight of the bulk atmosphere, or conversely if the weight is known the approximate temperature can be found. Rayleigh scattering is such an phenomenon and follows Equation~\ref{eq:sig_scale}, with $\alpha = -4$, and so the code was run with no other opacity sources in the wavelength region $\lambda \le 1.0 \mu m$ in order to validate the model. Performing this check, the code produced a slope which was within $2\%$ of the analytical Rayleigh slope (i.e. with $\alpha\,=\,-4$), which is in excellent agreement with the $1\%$ figure found by~\citet{Shabram+11}. 

However this treatment is not valid for opacity sources that do vary significantly with temperature and pressure, such as NIR absorption by molecules in general. A better test of the validity of the code is therefore obtained by noting that a recent study~\citep{Danielski+13} using the \textsc{tau} code was able to infer molecular abundances for the planet HD189733b that were in excellent agreement with the constraints published by other studies using a variety of different retrieval methods (see for example~\citet{Swain+08, Swain+09, Madhu+09}).

\section{Application}
\label{sec:application}
	The first step in the modelling procedure is to obtain the input files, in the correct formats. A P-T profile can be generated, ranging for example in pressure from $1.0 \times 10^{-3}\,Pa$ to $1.0 \times 10^{+6}\,Pa$ ($10\,nbar$ down to $10\,bar$). This range covers most of the relevant atmosphere, with corresponding altitudes calculated for each level using the hydrostatic assumption and the atmospheric scale height and planetary surface gravity (estimated with mass and radius values from the literature). The profile can in general be isothermal, with $T \approx T_{obs}$ of the planet, since transmission observations are currently not able to disentangle the degeneracy between the temperature profile and the absorber abundances, and so are insensitive to the exact form of the temperature profile. The absorption cross-section data for the relevant molecules must then be obtained, for example from the \textsc{hitemp}~\citep{Rothman+10} or \textit{Exomol}~\citep{Tennyson+12} line lists. Note that the data may need to be converted into different units - for example if absorption cross-sections are given in units of $cm^{2} g^{-1}$, they will need to be converted by multiplying each value by $\mu\,/\,N_{A}$, where $\mu$ is the mean molecular weight in atomic units (e.g. $\mu = 16$ for CH$_{4}$, $28$ for CO, $44$ for CO$_{2}$) and $N_{A}$ is Avogadro's constant. Absorption cross-sections should be calculated from line lists that are, to first order, at the same temperature as the planet $T \approx T_{obs}$, but can vary a little since the atmosphere is unlikely to be perfectly isothermal. The final file to be supplied is the data for the H$_{2}$-H$_{2}$ CIA effects, necessary when considering atmospheric pressures above approximately 1\,bar. These data can be taken from the literature (for example from~\citet{Borysow+01}), again converted to the specified format and correct units as appropriate. 

\begin{figure}
\caption{Spectra generated with no molecular absorbers, modifying the radius factors in order to calibrate the underlying H$_{2}$-H$_{2}$ continuum (`base level'). The system parameters were those for the planet HD189733b, as reported in~\citet{Schneider+11}, with an isothermal atmosphere at $1000K$, and the CIA file used for this was obtained from~\citet{Borysow+01} at $1000K$. Also note that Rayleigh scattering is evident here in the large slope at wavelengths $\lambda \ltapprox 0.5\,\mu m$.}
\label{fig:cia}
\smallskip
\scriptsize
\centering
	\includegraphics[width=0.8\columnwidth]{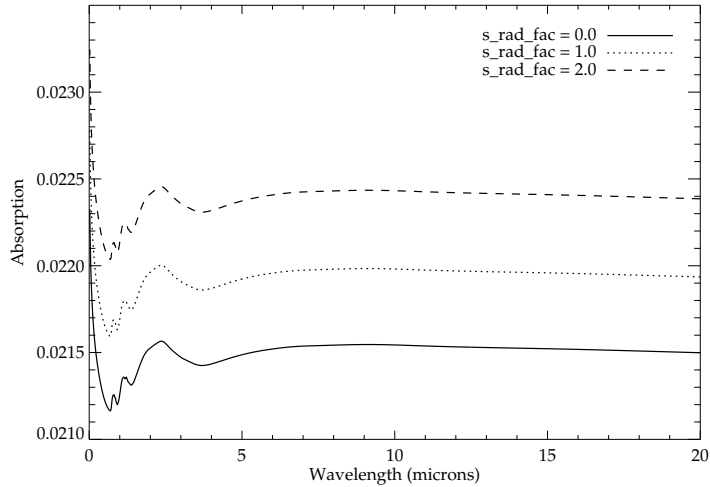}
\end{figure}

The next step is to calibrate the base level of the spectrum by changing the radius modifiers, in order to attempt to match the base level of any observations with all of the mixing ratios set to zero (i.e. no trace molecular absorbers present). The effect of this is to model the H$_{2}$-H$_{2}$ continuum, or minimum absorption level, on top of which the trace molecules will absorb, with output shown in Figure~\ref{fig:cia}. The underlying continuum will always be present when probing down to higher pressures, and allows effective constraints to be placed on the minimum absolute abundances of other trace absorber molecules if this is first calibrated to the highest possible value, and then the minimum possible amounts of other molecules are gradually added in. 

\begin{figure}
\caption{Adding H$_{2}$O into the model with a mixing ratio of $1.0\times10^{-5}$, using the zero offset Rayleigh and CIA baseline model from Figure~\ref{fig:cia}, and the absorption cross-sections used for water from the BT$2$ line list at $1000K$. The shape of the continuum can clearly still be seen, with the water absorption lines extending up from it into the higher atmosphere. \subref{subfig:h2o_zoom} shows the the spectrum in the NIR, with the H$_{2}$O absorption emerging from the CIA floor, and~\subref{subfig:h2o_full} shows the same model at a slightly lower model resolution on a larger scale from $0 - 20\,\mu m$.}
\label{fig:h2o}

\centering
\subfigure[]{
	\centering
	\includegraphics[width=0.8\columnwidth]{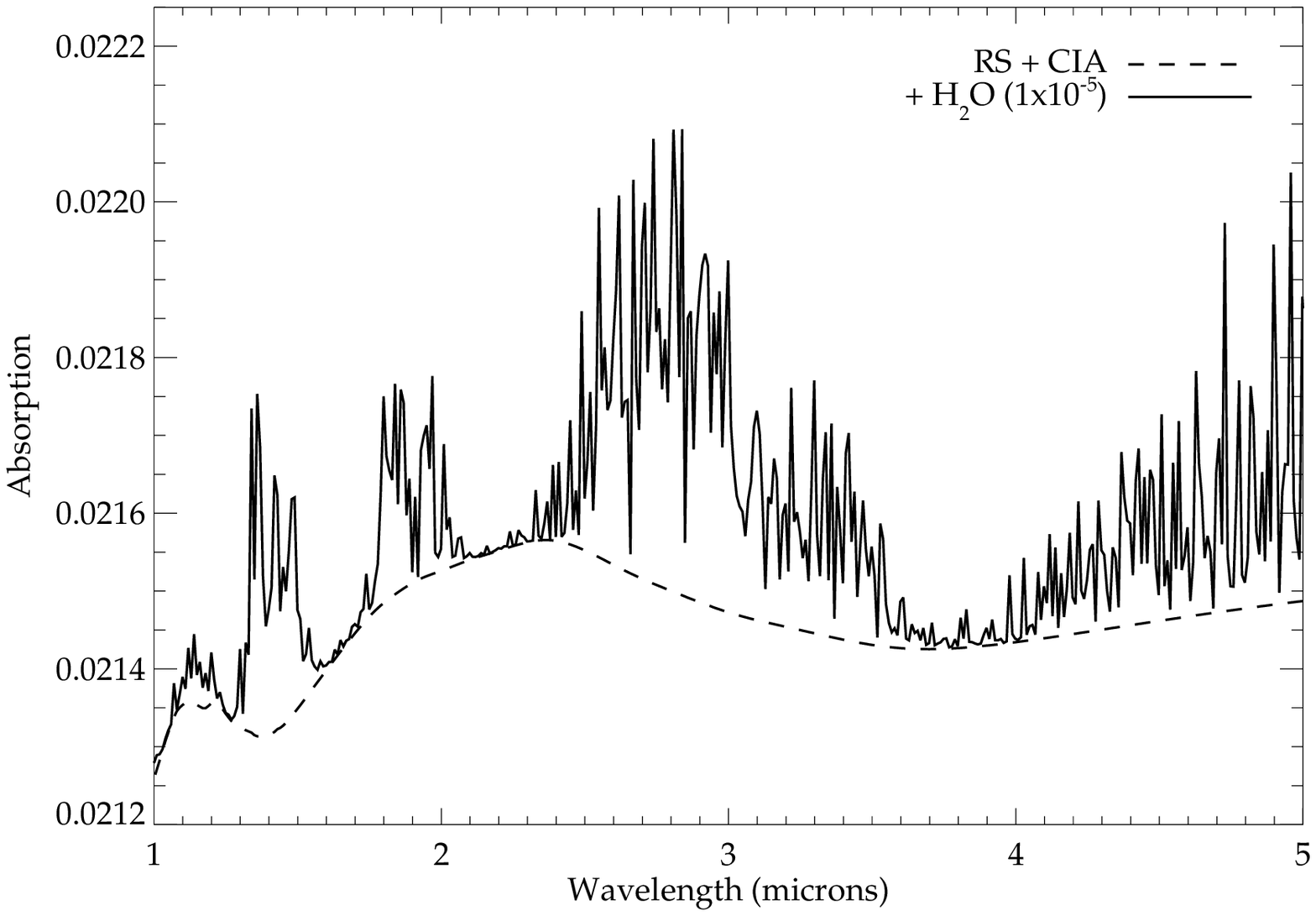}
	\label{subfig:h2o_zoom}
}
\subfigure[]{
	\centering
	\includegraphics[width=0.8\columnwidth]{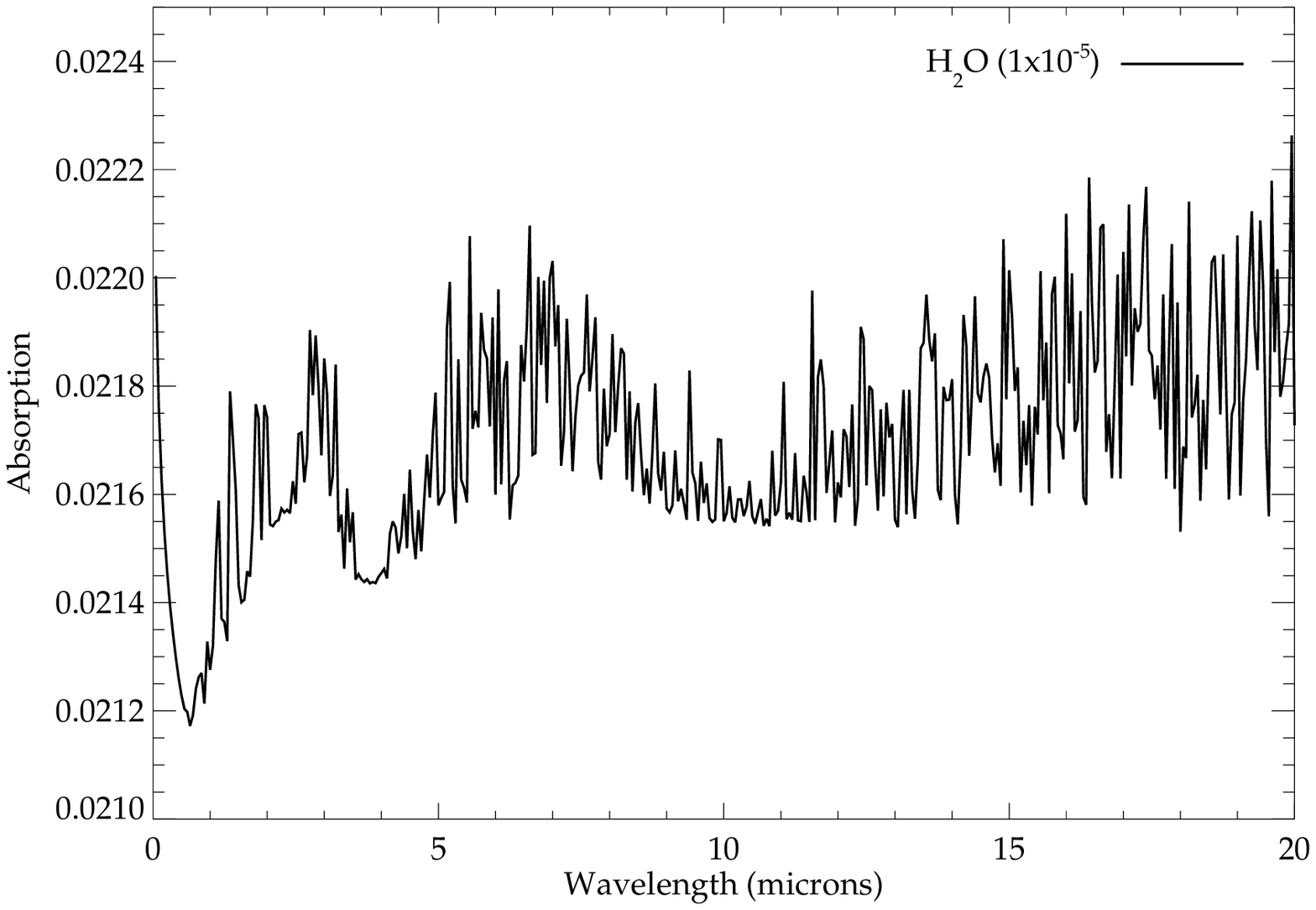}
	\label{subfig:h2o_full}
}

\end{figure}

The first trace molecule to consider adding into the model is H$_{2}$O, since it is found in relatively large abundances in known exoplanet atmospheres~\citep{Tinetti+12, Swain+09}, and is a very strong absorber in the NIR. In this range, water has characteristic absorption bands at wavelengths around $5.0 - 7.0\,\mu m$, with a narrower band around $3.0\,\mu m$, and steadily increasing broad band absorption as the wavelength increases (see Figure~\ref{fig:h2o}).

\begin{figure}
\caption{Contribution of CH$_4$ to the transmission spectrum, added on top of the base composition with a mixing ratio of $1.0\times10^{-4}$. Absorption cross-section data are from the \textit{Exomol} project, at $900K$.}
\label{fig:ch4}
\smallskip
\scriptsize
\centering
	\includegraphics[width=0.8\columnwidth]{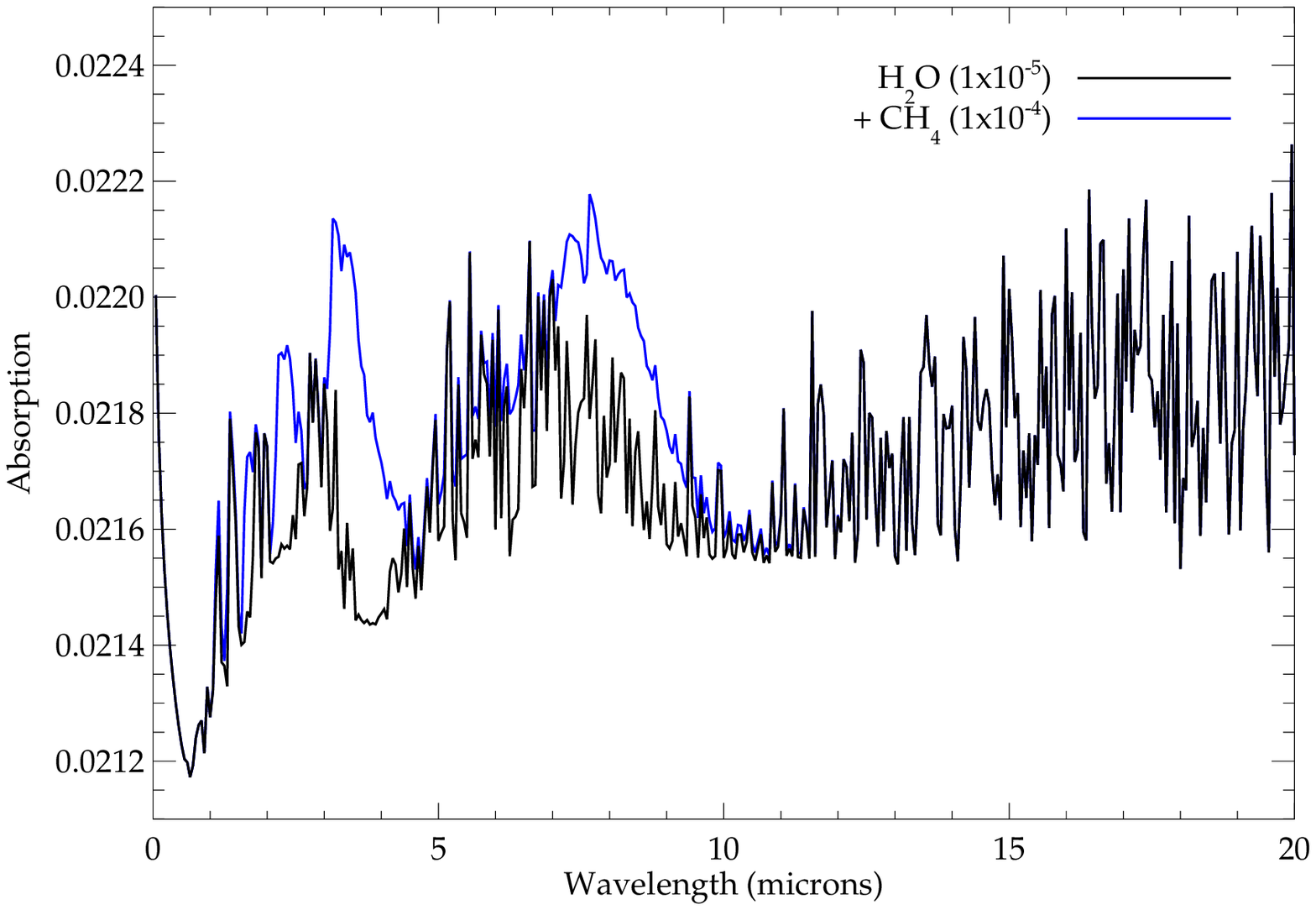}
\end{figure}

\begin{figure}
\caption{Contribution of CO to the transmission spectrum, added on top of the base composition with a mixing ratio of $1.0\times10^{-4}$. Absorption cross-section data are from \textsc{hitemp}, at $1000K$.}
\label{fig:co}
\smallskip
\scriptsize
\centering
	\includegraphics[width=0.8\columnwidth]{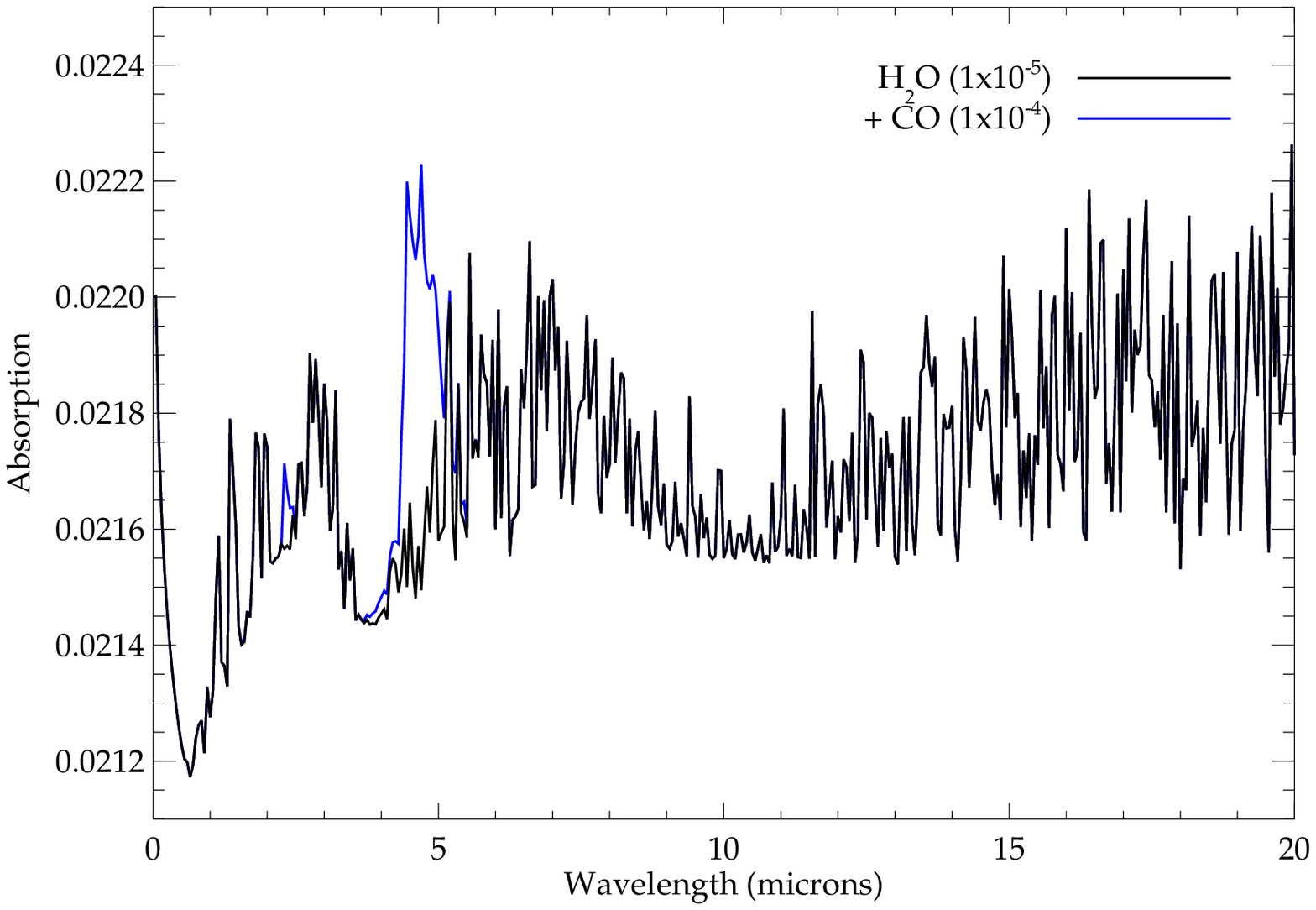}
\end{figure}

\begin{figure}
\caption{Contribution of CO$_{2}$ to the transmission spectrum, added on top of the base composition with a mixing ratio of $1.0\times10^{-5}$. Absorption cross-section data are from \textsc{hitemp}, at $1000K$.}
\label{fig:co2}
\smallskip
\scriptsize
\centering
	\includegraphics[width=0.8\columnwidth]{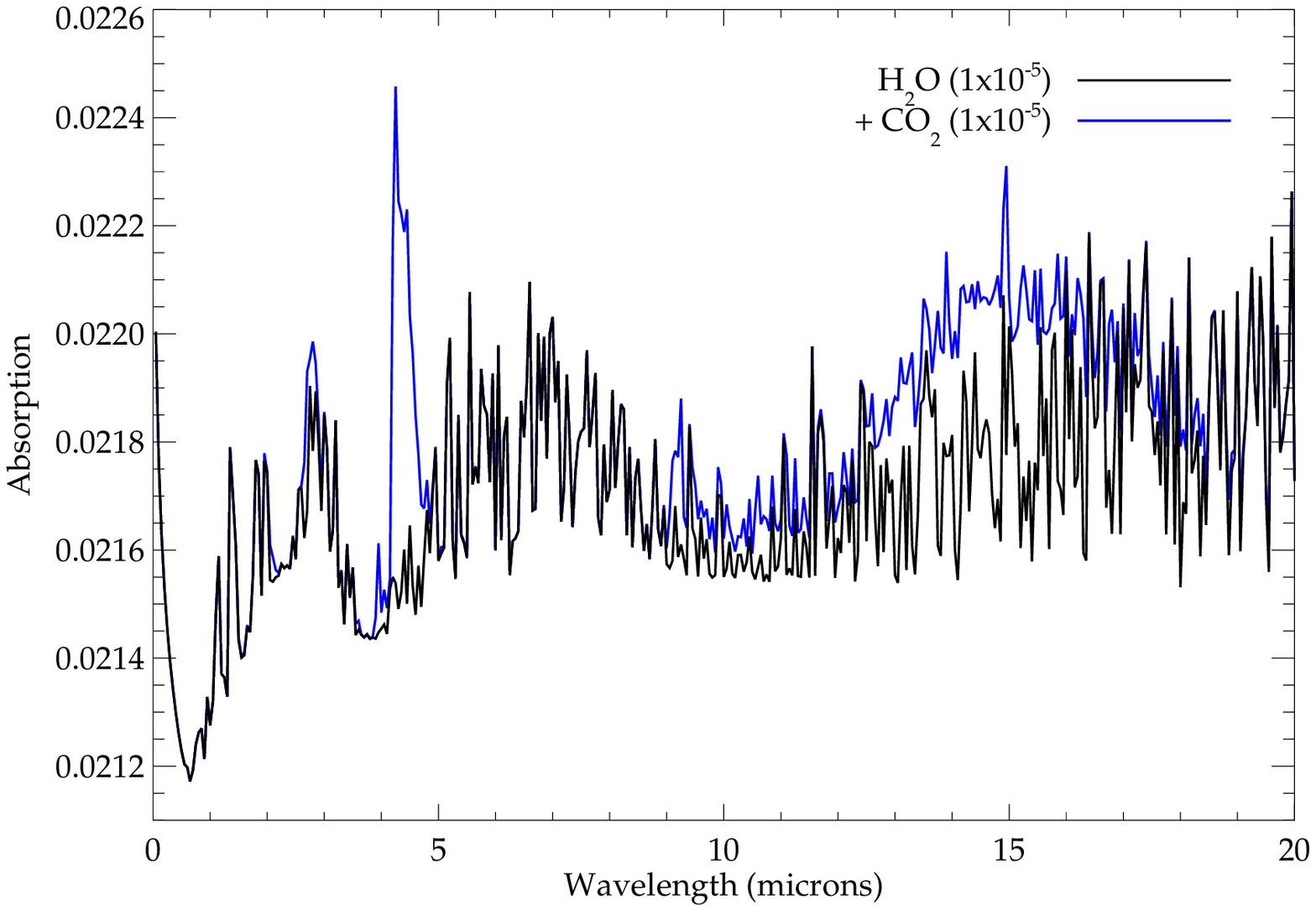}
\end{figure}

This process can then be repeated, adding in one constituent at a time, for as many molecules as are required (informed for example by results from atmospheric chemical equilibrium models). Other important absorbers include methane, which is also a strong NIR absorber and has broad characteristic features, and carbon monoxide and dioxide, which can also be present but with much narrower absorption peaks, except for CO$_2$ around $14.0 - 16.0\,\mu m$. If any extra opacities remain, further molecules may be added based on their typical absorption peaks and atmospheric chemistry considerations. Examples of the output from varying abundances of methane, carbon monoxide and carbon dioxide are shown in Figures~\ref{fig:ch4}, \ref{fig:co}, and~\ref{fig:co2}.

\section{Conclusion}
\label{sec:conclusion}
	This paper introduces the \textsc{tau} code, a $1D$ line-by-line radiative transfer code for transmission spectroscopy of extrasolar planets. It is generally applicable for the modelling of the transmission of radiation through the atmospheres of planets, in particular close-in gas giants. The result is a theoretical transmission spectrum of the planet, given the various inputs, which may be compared with relevant primary transit observations in order to match the spectra and hence infer characteristics of the atmosphere in question, such as composition and atmospheric structure. The code runs with the command \verb+./tau mode [input files]+ - it then proceeds to read in the assumed atmospheric P-T profile and absorption and scattering behaviour of the gases present, and interpolates to a defined wavelength grid. The optical depth as a function of wavelength is then calculated in a line-by-line integration scheme given input abundances for the trace constituents and a primary transit viewing geometry. 

The program, not being a chemical equilibrium code or GCM, does not model photochemistry~\citep{Moses+11}, dynamics, heat transfer or effects such as rainout~\citep{Burrows+99}, and assumes no time variation in the atmosphere. Studies using global circulation models indicate that these effects may be important and necessary in order to accurately and completely characterise a planetary atmosphere; such considerations are likely to be particularly pertinent when modelling the atmospheres of hot Jupiters (the most likely objects for which relevant data is available for fitting), since these objects are likely to be tidally locked, and hence possibly subject to significant temperature contrasts between day- and night-sides and extreme UV heating, with associated winds and large-scale circulation effects, and possible breakdown of hydrostatic and LTE assumptions. For more information on this subject the reader is referred to sources such as~\citet{Showman+09, Koskinen+07}. The code also does not yet have a complete treatment of cloud physics in terms of the detailed Mie scattering theory or geometric approximations at present, and so does not model such effects, though such a formulation will appear in a later version. However, codes modelling all of these effects are perfectly complementary to \textsc{tau}, and provide extra data that can improve the results. If such relevant data are available from other sources (e.g. absorption cross-sections for cloud types predicted to be present, calculated with an external program), these may easily be coupled to the \textsc{tau} code by simply providing the coefficient filenames and including them as extra opacities in the model. Similarly, results from other studies (e.g. dynamical and thermochemical models, in terms of modelling the atmospheric profile, which molecules are important etc.) can be used in conjunction with \textsc{tau} in order to inform the use of this code. In addition to the inclusion of more detailed cloud modelling, future work will involve streamlining the code, and treating all inputs self-consistently, for example calculating the atmospheric structure within the code and using all molecular absorption and collision data in the same format.

\section*{Acknowledgments}
\label{sec:acknowledge}
	MDJH is supported by an Impact/Perren studentship, and MT by an STFC studentship. GT is a Royal Society University Research Fellow. \\


\label{sec:refs}

\bibliographystyle{elsarticle/model1a-num-names}
\bibliography{tau_bib}

\appendix
\section{Opacities due to scattering}
\label{sec:app_scatter}
	Transmission through the atmosphere may additionally be decreased by extra absorption and scattering effects, such as Rayleigh scattering and collision-induced absorption (`CIA'). The size of a scattering particle can be expressed in the form of the \textit{size parameter}, $x$, such that,

\begin{equation}
\label{eq:size_param}
x \,=\, \frac{2 \pi a}{\lambda}, 
\end{equation}

for a particle of characteristic dimension $a$ and light of wavelength $\lambda$. Rayleigh scattering, the cause of the diffuse sky radiation that makes the sky appear blue, is caused by the scattering of electromagnetic radiation by particles much smaller than the wavelength of the radiation, i.e. in the small size parameter regime $x \ll 1$. Scattering from larger particles (applicable for example when considering the transmission of radiation through clouds, and when $x \gtrapprox 1$) is calculated using the geometric approximation or Lorenz-Mie-Debye theory, which reduces to the Rayleigh approximation for small $x$. Cloud scattering is not implemented in this version of the code, since the calculation of the Mie solutions depends in a complex manner on factors such as the particle size distribution in the cloud, and the precise shape of the scattering particles, and hence requires detailed cloud models (both photochemical and dynamical) and forward scattering models (see for example~\citet{deKok+12}), which are outside the scope of this project. 
	Collision-induced absorption on the other hand does not depend on the size parameter, but occurs in regions of the atmosphere at high pressure (i.e. the lower atmosphere), or along very long absorbing paths. Common diatomic molecules such as H$_2$, N$_2$ and O$_2$ do not normally absorb in the infra-red, since their symmetry means that they lack a permanent dipole moment and hence their rotational and vibrational energy levels cannot be excited by absorption of radiation. However, the collision processes that occur in such high pressure environments do induce a dipole moment, and hence these molecules suddenly become visible in absorption. In hot Jupiters the CIA due to hydrogen pairs is clearly the most important (since hydrogen is so abundant in these atmospheres) and a continuum is produced in the  transmission spectra, producing an effective absorption floor at this atmospheric level below which the infra-red radiation cannot penetrate.

\subsection{Rayleigh scattering}
\label{ssec:app_rayleigh}
	The optical depth due to Rayleigh scattering is given, for scattering particles at concentration $c \, = \, \chi \, \rho_{N}$ along a path $\int_{path} \, dl$, by,
\begin{eqnarray}
\label{eq:ray_tau}
	\tau_{R} (\lambda) & = & \sigma_{R} (\lambda) \, \int_{path} c \, dl \nonumber \\
	& = & \sigma_{R} (\lambda) \int_{path} \chi \, \rho_{N} \, dl ,
\end{eqnarray}

where $\chi$ is the mixing ratio and $\rho_{N}$ is the scattering particle number density. If the bulk atmosphere is composed of a mixture of species (e.g. H$_{2}$ - He mixes for hot Jupiters), the overall cross-section is simply the weighted sum over all of the constituents~\citep{Bates84}, and for the bulk atmosphere (i.e. with constant abundance profiles) the optical depth becomes, 
\begin{eqnarray}
\label{eq:ray_tau2}
	\tau_{R} & = & \sum_{i} \left[ \chi_{i} \, \sigma_{R,i} (\lambda) \right] \, . \, \int_{path} \rho_{N} \, dl \nonumber \\
	& \equiv & \sigma_{R,tot} (\lambda) \int_{path} \rho_{N} \, dl .
\end{eqnarray}

$\sigma_{R} (\lambda) $ is the Rayleigh scattering cross-section, given by~\citep{Liou02, Jackson99} as,
\begin{eqnarray}
\label{eq:ray_sigma}
	\sigma_{R} (\lambda) & = & \frac{ 128 \pi^{5} }{ 3 \lambda^{4} } \alpha^{2} \nonumber \\
	& = & \frac{128}{3} \frac{ \pi^{5} a^{6}}{\lambda^{4}}  (m_{r}^{2} - 1)^{2} f(\delta) ,
\end{eqnarray}

where $\alpha$ is the \textit{polarisability} of the particle, $m_{r}$ is the (real part of the) refractive index, $a$ is the scattering particle radius and $f(\delta)$ is a molecular anisotropy correction factor. Hence Rayleigh scattering is a function of the electric polarisability of the scattering particles, is independent of the density of the scattering medium, and has a wavelength dependence $\sim \lambda^{-4}$.

This equation is used in the function \verb$scatterRayleigh_H$, located in the header file. Since the Rayleigh cross-section $\sigma_{R}$ is a function only of the wavelength and particle properties and not of the atmospheric layer, computational expense can be saved by calling this function only once per bulk atmosphere species, at the start of the wavelength loop, with the wavelength $\lambda$ in $\mu m$ and the species class \verb+Mol+ as arguments. The function reads in, initialises and converts quantities as appropriate, before using the equation to calculate the species Rayleigh cross-section given those parameters, with the appropriate King correction factor for molecular asymmetry if necessary. This is returned and the final Rayleigh cross-section for the bulk atmosphere at that wavelength, $\sigma_{R,tot}(\lambda)$, is the weighted sum of the constituent species. For each atmospheric layer then, this total cross-section is multiplied by the layer optical path length, and hence the contributions of every layer summed to give the total Rayleigh optical depth at wavelength $\lambda$, as per Equation~\ref{eq:ray_tau2}.

As noted previously, this treatment of the Rayleigh scatter is only valid for small size parameters $x \ll 1$ - this condition is tested for in the main integration loop, with a cutoff defined arbitrarily as $\lambda \leq 50\,a$. If this is not valid in some regions of the defined wavelength range, a warning is printed and no Rayleigh scattering opacity is included in the overall sum for these wavelengths. This formulation is also subject to the assumption that absorption by the particles is negligible (i.e. the imaginary part of the refractive index is negligible and the particles are purely scattering), and further, that the refractive indices are constant with wavelength (though see for example~\citet{Dalgarno+62, Bates84}). This formulation is therefore only an approximate calculation of the Rayleigh scatter - although more complex aspects, such as refractive index changing with wavelength, temperature and pressure, are considered to be insignificant~\citep{Ehrenreich+06} and are not incorporated in this code, they would be required to treat this problem fully. The sample values for the refractive indices and particle sizes in the published code were taken from and calculated using the values and expressions in~\citet{Allen00}, using the Lorentz-Lorenz approximation for low density gases~\citep{Sneep+05}, and validated against the values and expressions from~\citet{Born33, Vardya62, desEtangs+08}.

\subsection{Collision-induced absorption}
\label{ssec:app_cia}
	CIA is calculated in the function \verb$scatterCIA_H$, called with the external absorption coefficients $\alpha$ in $cm^{-1} amg^{-2}$ (previously interpolated to the model wavelength grid in the function \verb$interpolateCS_H$) and atmospheric H$_{2}$ fraction. The return value is the absorption cross-section $\alpha_{cia} $ in $m^{5}mol^{-2}$. The full optical depth is dependent on the densities of both of the scattering particle types in every layer along the path as,
\begin{equation}
\label{eq:cia_tau1}
	\tau_{cia} = \int_{path} \, \alpha_{i,j} \, . \, c_{i} \, . \, c_{j} \, . \, dl ,
\end{equation}

for CIA from two particles $i$ and $j$. Hence for H$_{2}$-H$_{2}$ absorption, with $\chi_{i,j} \approx 1$, the optical depth is,
\begin{eqnarray}
\label{eq:cia_tau2}
	\tau_{cia} & = & \int_{path} \, \alpha_{i,j} \, . \, \rho_{i} \, . \, \rho_{j} \, . \, dl \nonumber \\
	& = & \int_{path} \, \alpha_{i,j} \, . \, \rho_{N}^{2} \, . \, dl .
\end{eqnarray}

Therefore within the function the read-in coefficients are converted from units of $cm^{-1} amg^{-2}$ to units of $cm^{5} mol^{-2}$ using the variable \verb$conv_factor$, which divides the coefficients by the square of Loschmidt's number $L_{0}$ (the number of molecules in one amagat at stp). The function then converts the coefficient to units of $m^{5}mol^{-2}$ and returns it, which gives a dimensionless value for $\tau_{cia}$, as required when multiplying the coefficients by $(mol \, m^{-3})^{2} \, . \, m$ in Equation~\ref{eq:cia_tau2}. However CIA coefficients for example from \textsc{hitran} (see~\citet{Richard+12}, already in units of $cm^{5} mol^{-2}$) can also be used by simply setting the value of \verb$conv_factor$ within the function to $1$.

\subsection{Cloud opacity estimation}
\label{ssec:app_cld}
	A basic estimation of the optical depth due to the presence of clouds is also included in this code, if files containing the relevant absorption coefficients are specified as inputs to the program. Following~\citet{Sharp+07}, if the cloud mass absorption coefficient $\alpha^{'}_{cld}(\lambda)$, in $cm^{2} g^{-1}$, is provided, then the cloud optical depth can be found by converting to the cloud absorption coefficient $\alpha_{cld}(\lambda)$, in $cm^{5} mol^{-2}$, with,
\begin{equation}
\label{eq:cld_tau}
	\tau_{cld} = \int_{path} \, \alpha_{cld} \, . \, \rho_{cld} \, . \, dl ,
\end{equation}

where $\rho_{cld}$ is the (mass) density of the cloud along the path. 

The density is found by interpolating between the specified pressure levels that define the vertical extent of the cloud, following the density profiles of~\citet{Ackerman+01} where $ln |\rho_{cld}| \, \sim \, ln |p|$. The units of $\rho_{cld}$ and $dl$ are then converted as appropriate for the absorption coefficients, and an estimate of the optical depth obtained as per Equation~\ref{eq:cld_tau}.

\end{document}